\documentclass[showpacs,floatfix,aps,prl,preprint]{revtex4}
\begin{document}
\title{Metrological constraints on the variability of the fundamental constants
 $e$, $\hbar$, and $c$}
\author{A.~Yu.~Ignatiev}
\email{a.ignatiev@physics.unimelb.edu.au}
\author{B.~J.~Carson}
\email{b.carson@physics.unimelb.edu.au} \affiliation{\em School of
Physics, Research Centre for High Energy Physics, The University of
Melbourne, Victoria 3010, Australia.}
\def\be{\begin{equation}}
\def\ee{\end{equation}}
\def\bea{\begin{eqnarray}}
\def\eea{\end{eqnarray}}
\newcommand{\nn}{\nonumber \\}
\begin{abstract}
 We set up a framework for a model-independent analysis of the time variation
  of $e$, $\hbar$, and $c$ individually. It is shown that  the
  time-evolution
   of each constant can be determined uniquely from the time evolution of
   the fine structure constant $\alpha$ provided that the choice of basic
   time-independent units (i.e., the clock
and ruler) is fixed. Realistic systems of units are considered as
examples and implications for metrology are discussed.
\end{abstract}
\maketitle

Recently evidence has been reported for a possible time evolution of the
fine structure constant on the cosmological time scale. The study of
absorbtion spectra of distant quasars yields $\Delta\alpha/\alpha\sim
10^{-4}$ over the redshift range $0.5<z<3.5$ \cite{webb1} \footnote{In
the latest study the range is $0.2<z<4.2$ \cite {mur}.}. Preliminary
results of a new study based on a larger sample support the earlier
results \cite{webb2}. However, a similar analysis performed by a
different group finds no evidence for the evolution of $\alpha$:
$\Delta\alpha/\alpha = (-0.06 \pm 0.06)\times 10^{-5}$ over the redshift
range $0.4<z<2.3$ \cite{sri}. Also, the variation of the fine structure
constant in $^{187}Re$ decay over the age of the solar system
(corresponding to $0<z\alt 0.4$) has been found consistent with zero at
the level of $\Delta\alpha/\alpha\sim 10^{-7}$ \cite{oli}. The source for
these discrepancies is yet to be understood.

The constancy of the proton-to-electron mass ratio $\mu=m_e/m_p$ is also
being investigated \cite{iva}.

Motivation for possible variability of fundamental constants appears
within different contexts such as Kaluza-Klein theories, superstring
theories and other models.  We will not go into theoretical discussion
   of what would be a microscopic model that could explain why
   fundamental constants change ( see e.g.
   \cite{uzan,mag,we,tob} and references therein \footnote{Mota and Barrow
    recently proposed the important idea of considering the
effect of inhomogeneity on the evolution of the fundamental constants
\cite{mb1,mb2,mb3}.}).
    Apart from specific
model building, a more general type of analysis is needed which would
start with kinematical (i.e., metrological) considerations. The role of
the metrological approach is to complement and clarify the specific
models (rather than to substitute them).

The question arises: if the fine structure constant does indeed evolve
with time then which of the constants $e$, $\hbar$, and $c$ are evolving
and how? This question has been at the centre of recent debate
\cite{duff1,duff2,mof,uzan,mag,nat,fla} and is the main topic of this
paper. Another issue of contention is the variation of dimensional
quantities, where on  one hand it is claimed that time variation of
dimensional constants is a meaningless concept. On the other hand, a
number of theories have been proposed in which one  (or more) dimensional
constants (such as $G$, $e$ or $c$) are  varying.

 Our analysis will explicitely show in what sense and to what extent the time evolution of dimensional constants is meaningful and when it becomes misleading; hence bringing the two points of view closer to each other and eliminating some controversy in this area.

 Fundamental constants and basic units form an interdependent system. This is
because the modern definitions of basic units (metre, second etc.) are
given in terms of processes controlled by the magnitudes of the
fundamental constants ($\alpha$, $\hbar$, $c$ etc.).
If fundamental constants would change with time then, generally, so would
the basic units.

Consequently, we face the following problem: how to define the basic
units in such a way that they stay invariant even though the fundamental
constants (which are involved  in their definitions) may change? This
problem should arise in any specific theory of $\alpha$ variation at the
stage when the contact with experiment is made; it is therefore desirable
to perform a model-independent analysis first.

The purpose of the this paper is to propose a possible solution to this
problem and to discuss some of its implications. In particular, we will
show that the time-dependence of $\hbar$, $c$, and $e$ can be
reconstructed uniquely from the time dependence of $\alpha$ provided that
the choice of basic units is fixed.

As a basis for our study we employ the standard methods of dimensional and
 metrological analysis supplemented by an additional postulate requiring
 that the fundamental units should be time-independent. In other words,
 we require that the units should be fixed even though constants can vary.
 Although it is possible, in principle, to use the system of time-dependent
 units that would greatly complicate both theory and experiment. Theoretically,
  one would have to keep track of time derivatives of units in all equations
  making them much longer; the number of experimental parameters to be
  measured would also greatly increase.

For simplicity and correspondence with the previous literature, the
centimetre-gram-second (CGS) system of units will be used. (This
assumption, however, is not important for the essence of our argument.)
Also, as there is no strong observational or experimental evidence for
the time-dependence of the fundamental constants apart from $\alpha$,
such as the mass ratios ($m_e/m_p$, $m_e/m_{Nucl}$, \dots) and
$g$-factors it will be assumed here that they do not vary with time (for
theoretical discussions see,e.g., \cite{cal,lan,dent}).

To start with, our units of time, length and mass can be defined in the
same way as the SI units (with proper scaling if necessary). Thus  $1 cm$
is the length travelled by light in a vacuum during a time interval of
1/299 792 45800 of a second, the second is the duration of 9 192 631 770
periods of the radiation corresponding to the transition between the two
hyperfine levels of the ground state of the cesium-133 atom and  $1 g$ is
1/1000 part of the mass of the international platinum-iridium prototype
of the kilogram.

These definitions imply that the magnitudes of 1 second ($s$), 1
centimetre ($\underline{cm}$) and 1 gram ($g$) depend on the magnitudes
of the fundamental constants as follows: \be\label{d} s \propto
\frac{1}{m_e\alpha^4}\frac{\hbar}{c^2},\ee \be\label{d20}  \underline{cm}
\propto c s \propto \frac{\hbar}{m_e\alpha^4 c},\ee \be\label{d1} g
\propto m_e. \ee
 It will be shown that the time evolution of $c$, $\hbar$ and $e$ crucially depend
  on the definitions of the centimetre and the second. To illustrate this point,
   several alternative definitions of the centimetre and second
   (see e.g. \cite{coh}) are introduced in the  following way.

A ``bar centimetre'' is 1/100 of the length of the international
platinum-iridium prototype of the metre. Its dependence on the
fundamental constants is controlled by the Bohr radius:
 \be\label{d2}  \underline{cm} \propto \frac{\hbar}{m_e\alpha c}\;\;\;
 (bar).\ee

 The ``krypton centimetre'' is the length equal to the 16507.6373 wavelengths
  of the radiation corresponding to the transition between the levels
   $2p_{10}$ and $5d_5$ of krypton-86; this length depends on the
    fundamental constants as the inverse Rydberg constant:

\be\label{e}\underline{cm} \propto \frac{\hbar}{m_e\alpha^2 c}\;\;\;
(krypton).\ee

(Both ``bar'' and ``krypton'' definitions of the metre were used by the
SI system in the past.)

 In addition, the second can be defined by choosing
among different types of clocks other than cesium-133 clock adopted by
the SI system; the ammonia clock (a representative of molecular clocks)
measures time in terms of molecular vibration frequencies \cite{coh}:
\begin{equation}
 \nu_{NH_3}\propto
\sqrt{ \frac{m_e}{m_p}}\frac{m_e\alpha^2 c^2}{\hbar}.
\end{equation}

An oscillator placed inside a superconducting cavity forms a
high-stability device called a SCSO clock. The frequency of such a device
is inversely proportional to the size of the superconductive cavity and
therefore to the Bohr radius \cite{coh}
\begin{equation}
\nu_{SCSO}\propto m_e\frac{\alpha c^2}{\hbar}.
\end{equation}
Correspondingly, the ``ammonia second'' depends on the fundamental
constants as \be\label{d10} s \propto \frac{\hbar}{m_e\alpha^2
c^2}\;\;\;(NH_3),\ee
 while the ``SCSO second'' has this dependence:
\be\label{f} s \propto \frac{\hbar}{m_e\alpha c^2}\;\;\;(SCSO).\ee Thus
the three different definitions of the second are described by
\be\label{n} s_n \propto \frac{\hbar}{m_e\alpha^n c^2}, \ee where
$n=1,2,4$ corresponds to the SCSO, ammonia and
 cesium clock, respectively \cite{fn}.

Similarly, the three different definitions of the centimetre can be
combined as \be\label{l} \underline{cm}_l \propto
\frac{\hbar}{m_e\alpha^l c} \ee with $l=1, 2, 4$ for the bar, krypton and
light definitions of unit of length, correspondingly.

From dimensional considerations it follows that an arbitrary system of
units, $S_{nl}$, can be characterized by Eq.(\ref{n}) and (\ref{l}) with a
suitable choice of clock and ruler indices $n$ and $l$. While
theoretically all these $S_{nl}$ systems are equivalent, only a small
number of them can be of interest for the past and future high-precision
experiments. The nine systems considered in this paper form a
representative sample of such realistic systems of units.

Requiring that the centimetre, second and gram do not depend on time, we
obtain \bea  c & \propto \alpha^{l-n} \label{main}\\  \hbar & \propto
\alpha^{2l-n}\\
 e & \propto \alpha^{\frac{1+3l-2n}{2}}. \label{main1}\eea

These can be conveniently rewritten as \bea\label{g}\frac{\Delta
c}{c}&\simeq (l-n)\frac{\Delta \alpha}{\alpha} \\ \frac{\Delta
\hbar}{\hbar}& \simeq (2l-n)\frac{\Delta \alpha}{\alpha} \\  \frac{\Delta
e}{e}& \simeq
 \left(\frac{1+3l-2n}{2}\right)\frac{\Delta
\alpha}{\alpha}.\eea

Thus, if time dependence of $\alpha$ is experimentally measured {\em and}
units are fixed, there is {\em no} further choice on how $e, c, \hbar$
depend on time and no extra measurements are required. The results are
presented in Table \ref{tb:clocks&rulers}.
\begin{table}
\begin{center}
\begin{tabular}{|c|c|c|c|}\hline
            & $n=1$       & $n=2$        & $n=4$     \\ \hline
            &             &              &           \\
            & $\frac{\Delta c}{c}=0$   & $\frac{\Delta c}{c}= -\frac{\Delta \alpha}{\alpha}$  & $\frac{\Delta c}{c}= -3\frac{\Delta \alpha}{\alpha}$\\
$l=1$       & $\frac{\Delta\hbar}{\hbar}=\frac{\Delta \alpha}{\alpha}$ & $\frac{\Delta\hbar}{\hbar}=0$ & $\frac{\Delta\hbar}{\hbar}= -2\frac{\Delta \alpha}{\alpha}$\\
            & $\frac{\Delta e}{e}=2\frac{\Delta \alpha}{\alpha}$ & $\frac{\Delta e}{e}=0$ & $\frac{\Delta e}{e}=-2\frac{\Delta \alpha}{\alpha}$\\
            &             &              &          \\ \hline
            &             &              &           \\
            & $\frac{\Delta c}{c}=\frac{\Delta \alpha}{\alpha}$ & $\frac{\Delta c}{c}=0$ & $\frac{\Delta c}{c}= -2\frac{\Delta \alpha}{\alpha}$\\
 $l=2$      & $\frac{\Delta\hbar}{\hbar}=3\frac{\Delta \alpha}{\alpha}$ & $\frac{\Delta\hbar}{\hbar}=2\frac{\Delta \alpha}{\alpha}$ & $\frac{\Delta\hbar}{\hbar}= 0$\\
            & $\frac{\Delta e}{e}=2.5\frac{\Delta \alpha}{\alpha}$ & $\frac{\Delta e}{e}=1.5\frac{\Delta \alpha}{\alpha}$ & $\frac{\Delta e}{e}=-0.5\frac{\Delta \alpha}{\alpha}$\\
            &             &              &           \\ \hline
            &             &              &           \\
            & $\frac{\Delta c}{c}=3\frac{\Delta \alpha}{\alpha}$ & $\frac{\Delta c}{c}=2\frac{\Delta \alpha}{\alpha}$ & $\frac{\Delta c}{c}=0$\\
 $l=4$      & $\frac{\Delta\hbar}{\hbar}=7\frac{\Delta \alpha}{\alpha}$ & $\frac{\Delta\hbar}{\hbar}=6\frac{\Delta \alpha}{\alpha}$ & $\frac{\Delta\hbar}{\hbar}= 4\frac{\Delta \alpha}{\alpha}$\\
            & $\frac{\Delta e}{e}=5.5\frac{\Delta \alpha}{\alpha}$ & $\frac{\Delta e}{e}=4.5\frac{\Delta \alpha}{\alpha}$ & $\frac{\Delta e}{e}=2.5\frac{\Delta \alpha}{\alpha}$\\
            &             &              &           \\ \hline
\end{tabular}
\vspace{4mm} \caption{Variation of the fundamental constants as measured
by different clocks, $n$ (Eq. \ref{n}), and rulers, $l$ (Eq. \ref{l}). }
\label{tb:clocks&rulers}
\end{center}
\end{table}

How many dimensional constants vary? In all cases except one, at least
{\em two} constants must vary. These varying constants {\em cannot} be
set equal to 1 and treated as mere conversion factors. The exceptional
case is $l=1,n=2$, i.e. the ``bar'' ruler plus the ammonia clock. In this
case $c\propto 1/\alpha$ while $e$ and $\hbar$ stay constant, which
allows one to use the system $e=1$, $\hbar=1$ (Hartree units). In other
words, in this case $e$ and $\hbar$ {\em can} be treated as mere
conversion factors.

Is the speed of light constant?  For the case of the ``light centimetre''
(i.e., SI-like definition), $c$ remains constant automatically; however,
 in general, $c$ varies. Interestingly, for any choice of the centimetre
 there exists a "matching choice" of the second (i.e., $n=l$) so that
 the speed of light remains constant. Furthermore, if we adopt a system
 with
 $l=n$  (in such systems $c$ is constant) then it follows that a) {\em both} $e$
  and $\hbar$ were smaller in the past, and b) the SCSO clock minimizes
  the time dependence of $\hbar$ and $e$ (because this clock has lower $n$
   than other clocks).

In  theoretical papers aimed at explaining the $\alpha$ variation in  the
context of quantum field theory the choice $\hbar=c=1$ is frequently
made. Consequently, it is implicitely assumed that both $\hbar$ and $c$
do not depend on time. From Eq.(15) and (16) it follows then that the
only possible choice for $n$ and $l$ is $n=0$ and $l=0$, i.e. the system
of units must be of $S_{00}$-type. This system differs from the systems
of units used in actual experiments (such as systems in Table 1 -- e.g.,
the SI system is of $S_{44}$-type). In other words, the theorist's units
would appear time-dependent from the point of view of the experimenter
(and the other way round). Therefore the translation from one system to
the other should be done explicitely in order to make the contact between
the theory and observations.

An interesting question is whether or not one can design a
 high-precision clock with $n=0$ or $n=-1$. Such  clocks would be
  useful in removing the time dependence of $c$ and $\hbar$, or $c$
  and $e$ respectively.

These results should be taken into account while comparing the
predictions of $\alpha$ varying theories and the forthcoming experiments
on high-precision clock comparison.
Particular attention should be taken to ensure that the
theoretical and experimental results are
 expressed in {\it the same} system  of units. In the case that experiment
 finds a discrepancy between the readings of two different clocks, and thus
  confirms the variability of $\alpha$, the problem will arise as to which
  clock shows the ``true'' time. Our discussion would help in making this
   choice.

A topic of active debate in recent literature has been whether or not it
is meaningful to talk about the time evolution of {\em dimensional}
fundamental constants. The discussion presented in this paper goes some
way in resolving this problem by demonstrating a framework in which time
dependence of {\em dimensional} constants becomes a well defined concept.
Such frameworks require a set of basic assumptions to be fixed (in this
particular case, that the units of length, time and mass remain
constant.) This can be compared to gauge-dependent quantities and gauge
fixing in quantum field theory.

 In
summary, the time evolution of the three fundamental constants $e$,
$\hbar$ , and $c$ has been studied from a metrological perspective.
Assuming that
 $\alpha$ evolution is known from observation, and the choice of time-independent
 units is
 fixed, it has been shown that the three separate evolution laws of $e$,
  $\hbar$, and $c$ can be found explicitely for a generic system of units.
   As an application of these findings, nine specific unit systems of
    interest for experiment and theory, based on different realistic choices
     of clocks and rulers, have been considered. Time variations of $e$, $\hbar$,
      and $c$ in these nine cases are found and compared, and their
      significance to the forthcoming experimental tests of $\alpha$ variation
      are discussed.

We are grateful to G.C.Joshi, W.McBride, B.H.J.McKellar and R.R.Volkas
for stimulating discussions and to T.Dent and D.F.Mota for valuable
comments.


\begin{thebibliography}
\bibitem{fn}These equations are
valid neglecting higher order corrections. Inclusion of these corrections
will lead to effective non-integer values of n:$n\rightarrow
n_{eff}=n+\delta_n$.
\end{thebibliography}
\end{document}